\documentstyle[preprint,eqsecnum,aps]{revtex}
\tighten

\begin{document}
\draft

\preprint{SLAC-PUB-6661}

\title{
Beta decay of hyperons in a relativistic quark model
}
\author{Felix Schlumpf}
\address{
Stanford Linear Accelerator Center\\
Stanford University, Stanford, California 94309
}
\date{\today}
\maketitle

\begin{abstract}
A relativistic constituent quark model is used to calculate the
semileptonic beta decay of nucleons and hyperons. The parameters of the
model, namely, the constituent quark mass and the confinement scale, are
fixed by a previous calculation of the magnetic moments of the baryon octet
within the same model. We discuss the momentum dependence of the form
factors, possible configuration mixing and SU(3) symmetry breaking. We
conclude that the relativistic constituent quark model is a good framework
to analyze electroweak properties of the baryons.

\end{abstract}


\narrowtext

\section{Introduction}\label{sec:1}

In this paper we consider the application of the relativistic constituent
quark model to the semileptonic hyperon decay. We compare our result with
the new data from the particle data group~\cite{part94}.
The predictive power of a relativistic
constituent quark model formulated on the light-front was recently
investigated in Ref.~\cite{schl93a}. It
provides a simple model wherein we have overall an excellent and
consistent picture of the magnetic moments and of the semileptonic decays
of the baryon octet. This paper extends the analysis of the semileptonic
beta decays and addresses specific questions for the hyperon beta decay.

The effect of configuration mixing has recently been studied~\cite{dzie94}
in the context of deep inelastic scattering. We show below that such a
configuration mixing is not favored for hyperon decays.

Our quark model provides a unique scheme for
calculating the momentum dependence of the form factors.
Although it is generally small, a change of the dipole masses $M_V$ or
$M_A$ by $\pm 0.15$ GeV in the case $\Sigma^- \to ne\nu$
causes a relative change of $g_1/f_1$ of
$\pm 2\%$. Ignoring the momentum dependence altogether would shift
$g_1/f_1$ by $17 \%$.

The SU(3) symmetry breaking can also be studied in our model. It plays a
major role in the determination of the Kobayashi-Maskawa-Cabibbo matrix
element $V_{us}$ from baryon decay.

The parameters of the model are the constituent quark mass $m$ and the
scale parameter $\beta$, which is a measure for the size of the baryon.
All parameters have been determined and fixed in Ref.~\cite{schl93a}.
The results reported in this paper are independent of the wave function
assumed in the calculation. It has been shown in Ref.~\cite{schl94b} that
relations between observables at zero momentum transfer are independent of
the wave function, and Ref.~\cite{schl94a} shows that this independence
holds up to 1 GeV$^2$ for the baryons.

This article is organized as follows.
Section~\ref{sec:2} describes the basics of hyperon semileptonic decay. In
Sec.~\ref{sec:3} we give a brief summary of our model as described in
Ref.~\cite{schl93a} with the explicit expressions for the beta decay. The
numerical results are presented in Sec.~\ref{sec:4}, and are compared with
experiment, other calculations, and some extensions of the model. We
summarize our investigation in a concluding Sec.~\ref{sec:5}.

\section{Hyperon semileptonic decay}\label{sec:2}

In the low energy limit the standard model for semileptonic weak decays
reduces to an effective current-current interaction Hamiltonian
\begin{equation}
	H_{\text{int}} = {G \over \sqrt{2}} J_\mu L^\mu + \text{h.c. }\;,
\end{equation}
where $G \simeq 10^{-5}/M_p^2$ is the weak coupling constant,
\begin{equation}
L^\mu = \bar{\psi_e} \gamma^\mu (1 - \gamma_5) \psi_\nu +
   \bar{\psi_\mu} \gamma^\mu (1 - \gamma_5) \psi_\nu
\end{equation}
is the lepton current, and
\begin{eqnarray}
J_\mu & = & V_\mu - A_\mu\;,
	\nonumber \\
V_\mu & = & V_{ud} \bar u \gamma_\mu d + V_{us} \bar u \gamma_\mu s\;,
	 \\
A_\mu & = & V_{ud} \bar u \gamma_\mu \gamma_5 d + V_{us} \bar u \gamma_\mu
\gamma_5 s \;,
	\nonumber
\end{eqnarray}
is the hadronic current, and $V_{ud}, V_{us}$ are the elements of the
Kobayashi-Maskawa mixing matrix.
The $\tau$-lepton current cannot
contribute since $m_\tau$ is much too large.

The matrix elements of the hadronic current between spin-${1 \over 2}$
states are
\begin{equation}
\left< B',p' \left| V^\mu\right| B,p \right> = V_{qq'} \bar u(p') \left[
  f_1(K^2) \gamma^\mu - {f_2(K^2) \over M_i} i \sigma^{\mu\nu} K_\nu
  + {f_3(K^2) \over M_i}  K^\mu \right] u(p)\;,
\end{equation}
\begin{equation}
\left< B',p' \left| A^\mu\right| B,p \right> = V_{qq'} \bar u(p') \left[
  g_1(K^2) \gamma^\mu - {g_2(K^2) \over M_i} i \sigma^{\mu\nu} K_\nu
  + {g_3(K^2) \over M_i}  K^\mu \right] \gamma_5 u(p)\;,
\end{equation}
where $K = p - p'$ and $M_i$ is the mass of the initial baryon.
The quantities $f_1$ and $g_1$ are the vector and axial-vector
form factors, $f_2$ and $g_2$ are the weak magnetism and electric form
factors and $f_3$ and $g_3$ are the induced scalar and pseudoscalar form
factors, respectively. Time invariance implies real form factors.
We do not calculate $f_3$ and $g_3$ since we put $K^+ = 0$
and their dependence on the decay spectra is of the order
\begin{equation}
\left( m_l \over M_i \right)^2 \ll 1\;,
\end{equation}
where $m_l$ is the mass of the final charged lepton.
The other form factors are
\begin{eqnarray}
f_1 &=&\left< B',\uparrow\left| V^+\right| B,\uparrow \right>
\;,\nonumber\\
K_\perp f_2 &=&M_i\left< B',\uparrow\left| V^+\right|
B,\downarrow \right>\;,\nonumber\\
g_1 &=&\left< B',\uparrow \left| A^+\right| B,\uparrow\right>
\;,\nonumber\\
K_\perp g_2 &=&-M_i\left< B',\uparrow \left| A^+\right| B,
\downarrow\right>\;.
\label{eq:formfactors}
\end{eqnarray}

What is usually measured is the total decay rate $\Gamma$,
the electron-neutrino
correlation $\alpha_{e\nu}$ and the electron $\alpha_e$, neutrino
$\alpha_\nu$ and final baryon $\alpha_B$ asymmetries.
The $e$-$\nu$ correlation is defined as
\begin{equation}
\alpha_{e\nu} = 2{N(\Theta_{e\nu} < {1 \over 2}\pi) - N(\Theta_{e\nu} >
 {1 \over 2}\pi) \over N(\Theta_{e\nu} < {1 \over 2}\pi) +
N(\Theta_{e\nu} > {1 \over 2}\pi)}\;,
\end{equation}
where $N(\Theta_{e\nu} < {1 \over 2}\pi)$ is the number of $e$-$\nu$
pairs that form an angle $\Theta_{e\nu}$ smaller than $90^\circ$.
The correlations $\alpha_e$,$\alpha_\nu$ and $\alpha_B$ are defined
analogously with
$\Theta_e$,$\Theta_\nu$ and $\Theta_B$ now being the angles between
the $e$, $\nu$, $B$ directions and the polarization of the initial
baryon.

Ignoring the lepton-mass one can calculate expressions for the
measured quantities. Expressions for $\Gamma$, $\alpha_{e\nu}$,
$\alpha_e$, $\alpha_\nu$ and $\alpha_B$ are given in
Ref. \cite{garc85}. For the decay rate $\Gamma$ we have for instance:
\begin{eqnarray}
\Gamma &=&G^2 {\Delta M^5 |V|^2 \over 60\pi^3}\Bigl[ (1 - {3 \over 2} \beta +
  {6 \over 7} \beta^2) f_1^2 + {4 \over 7} \beta^2 f_2^2 + (3 -
  {9 \over 2} \beta + {12 \over 7} \beta^2) g_1^2  \nonumber\\
  &&+ {12 \over 7} \beta^2 g_2^2
  + {6 \over 7} \beta^2 f_1 f_2 + (-4\beta + 6\beta^2) g_1 g_2
  + {4 \over 7} \beta^2 (f_1 \lambda_f + 5g_1\lambda_g) \Bigr]\;,
\label{eq:rate}
\end{eqnarray}
where $\beta$ is defined as $\beta = (M_i - M_f) / M_i$,
and $\Delta M =M_i-M_f$,
$M_i$, $M_f$ being the masses of the initial and final baryon,
respectively. The $K^2$ dependence of $f_2$ and $g_2$ is ignored and
$f_1$ and $g_1$ are expanded as
\begin{equation}
f_1(K^2) = f_1(0) + {K^2 \over M_i^2} \lambda_f\;, \quad
g_1(K^2) = g_1(0) + {K^2 \over M_i^2} \lambda_g \;.
\end{equation}
We get the corresponding expression for the dipole parameterization
$f(K^2)=(1-K^2/M^2)^{-2}$ by putting
\begin{equation}
\lambda_f = 2 M_i^2 f_1 / M_V^2\;, \quad
\lambda_g = 2 M_i^2 g_1 / M_A^2 \;.
\end{equation}

These quantities are corrected by the nonvanishing lepton mass and
radiative corrections~\cite{garc85,garc82,gail84}.

\section{The form factors in a relativistic constituent quark model}
\label{sec:3}
The constituent quark model described in Ref.~\cite{schl93a} provides a
framework for representing the general structure of the three-quark wave
function for baryons. The model is formulated on the light-front, which is
specified by the invariant hypersurface $x^+ = x^0 + x^3 = 0$.
The wave function is constructed as the product of a
momentum wave function, which is
spherically symmetric and invariant under permutations,
and a spin-isospin wave function,
which is uniquely determined by SU(6) symmetry requirements.  A
Wigner (Melosh) rotation~\cite{melo74}
is applied to the spinors,
so that the wave function of the proton is an eigenfunction
of $J^2$ and $J_z$ in
its rest frame~\cite{coes82}. To represent the range of uncertainty in
the possible form of the momentum wave function, harmonic oscillator
and a pole-type wave function have been chosen in
Refs.~\cite{schl93a,schl94b,schl94a}. Surprisingly, it has been
found that observables at zero momentum transfer are independent of the
wave function chosen~\cite{schl94b}, and form factors do not differ
up to 1~GeV$^2$~\cite{schl94a} for a wide range of wave functions.
Since the momentum transfer involved in hyperon beta decays is much
smaller than 1~GeV$^2$ it is representative to use one special wave function.
The form factors in Eq.~\ref {eq:formfactors} are calculated as shown in
Ref.~\cite{schl93a}. In contrast to Ref.~\cite{schl93a}, we do not
assume additional structure of the constituent quarks, and we choose
symmetric wave functions. These simplifications reduce the number
of free parameters to two masses ($m_{u/d}, m_s$) and three
scale parameters ($\beta_N, \beta_{\Sigma /\Lambda}, \beta_\Xi$).

For $K^2 =0$ we have for $\Delta S = 0$ transitions
\begin{eqnarray}
f_1 &=&A(f_1) \;,\nonumber\\
f_2 &=&{N_c \over (2\pi)^6}\int d^3q d^3Q |\Phi|^2 A(f_2) \;,\nonumber\\
&&\label{eq:5.20}  \\
g_1 &=&A(g_1) {N_c \over (2\pi)^6} \int d^3q d^3Q |\Phi|^2
{b^2 - Q_\bot^2 \over b^2 + Q_\bot^2} \;,\nonumber\\
g_2 &\simeq &0\;, \nonumber
\end{eqnarray}
with $A$s given in Table \ref{tab:reaction1}.
The values $A(f_1)$ and $A(g_1)$ are the values in the
nonrelativistic quark model. The factors $A_1$, $A_2$, and $A_3$ are
given in Eq.~(3.6) of Ref.~\cite{schl93b}.

The $\Delta S = 1$ transitions for $K^2 = 0$ are
\begin{eqnarray}
f_1 &=&{N_c \over (2\pi)^6}\int d^3\!q d^3Q \left({E_3'E_{12}'M \over
E_3 E_{12} M'} \right)^{1/2}\!{\Phi^\dagger(M')\Phi(M)
B(f_1) \over (a'^2+Q_\bot^2)(a^2+Q_\bot^2)\sqrt{b'^2+Q_\bot^2}
\sqrt{b^2+Q_\bot^2}}\;,\nonumber\\
&&\label{eq:weak1}\\
g_1 &=&{N_c \over (2\pi)^6}\int d^3\!q d^3Q \left({E_3'E_{12}'M \over
E_3 E_{12} M'} \right)^{1/2}\!{\Phi^\dagger(M')\Phi(M)
B(g_1) \over (a'^2+Q_\bot^2)(a^2+Q_\bot^2)\sqrt{b'^2+Q_\bot^2}
\sqrt{b^2+Q_\bot^2}} \;,\nonumber
\end{eqnarray}

\begin{eqnarray}
B(f_1) &=&B_1(a'a+Q_\bot^2)^2(b'b+Q_\bot^2)\nonumber\\
 &&+B_2(a'-a)^2Q_\bot^2(b'b+Q_\bot^2){(cd-q_\bot^2)^2 \over (c^2+q_\bot^2)
(d^2+q_\bot^2)}\nonumber\\
 &&+B_3(a'-a)(b'-b)Q_\bot^2(a'a+Q_\bot^2)\left({c^2 \over c^2+q_\bot^2}
+ {d^2 \over d^2+q_\bot^2} \right)\;,\nonumber\\
&&\label{eq:5.22}\\
B(g_1) &=&B_4(b'b-Q_\bot^2)\left[(a'a+Q_\bot^2)^2 + (a'-a)^2Q_\bot^2
{(cd-q_\bot^2)^2 \over (c^2+q_\bot^2)(d^2+q_\bot^2)} \right]\nonumber\\
 &&+B_5(a'-a)^2Q_\bot^2(b'b-Q_\bot^2){cdq_\bot^2 \over (c^2+q_\bot^2)
(d^2+q_\bot^2)}\nonumber\\
 &&+B_6(a'-a)Q_\bot^2(b'+b)(a'a+Q_\bot^2)\left({c^2 \over c^2+q_\bot^2}
+ {d^2 \over d^2+q_\bot^2} \right)\;.\nonumber
\end{eqnarray}
The $B_i$ for the different decays are given in Table \ref{tab:reaction2}.
The quantities $a, b, c, d, q_\perp, Q_\perp, E_3, E_{12}$, and $M$ are
defined in Eqs.~(2.1), (2.3), (2.4) and (3.8) of Ref.~\cite{schl93a}.

Eqs.~(\ref{eq:weak1}) and (\ref{eq:5.22}) confirm the Ademollo-Gatto theorem
\cite{adem64}.
Since $(a'-a) \sim \Delta m$ and $(b'-b) \sim \Delta m$ the symmetry
breaking for $f_1$ is of the order $(\Delta m)^2$ whereas it is of the order
$\Delta m$ for $g_1$ owing to the term containing $B_6$. In addition
to Ademollo-Gatto
we see that the symmetry breaking for $g_1(\Lambda \to p)$ is
also of second order.

The full formulae for $K^2 \le 0$ are longer than the ones for
$K^2 = 0$; they are given in Ref.~\cite{schl92}.

\section{Numerical results}\label{sec:4}

The form factors can be determined by the generalization of
Eqs.~(\ref{eq:5.20}) and (\ref{eq:weak1}). With the parameterization
of the form factor $f(K^2)$:
\begin{equation}
f(K^2) \simeq {f(0) \over 1-K^2 / \Lambda_1^2+K^4 / \Lambda_2^4} \;,
\label{eq:pol}
\end{equation}
we get the result shown in Tables \ref{tab:nstrange} and
\ref{tab:strange} together
with the rates, angular correlation and asymmetries.
The parameters $\Lambda_n$ are determined by the
calculation of the appropriate derivatives of $f(K^2)$ at
$K^2=0$. The rates have been corrected taking into account
the nonvanishing lepton mass and radiative corrections.

In this paper, we use the parameter set~2 of Ref~\cite{schl93a}. The
values for the constituent quark masses and the confinement scales are
\begin{eqnarray*}
	m_u=m_d & = & 0.267 \text{ GeV } ,
	 \\
	m_s & = & 0.40 \text{ GeV } ,
	 \\
	\beta_N & = & 0.56 \text{ GeV } ,
	 \\
	\beta_\Sigma = \beta_\Lambda & = & 0.60 \text{ GeV } ,
	 \\
	\beta_\Xi & = & 0.62 \text{ GeV } .
\end{eqnarray*}
These parameters also give good results for the magnetic moments of the
baryon octet~\cite{schl93a}.

\subsection{The rates, $f_1(0)$, and $g_1(0)$}

The largest discrepancy between theory and experiments comes from the rates
and $g_1/f_1$ for the processes $\Lambda\to pe^-\bar\nu_e$ and
$\Sigma^-\to ne^-\bar\nu_e$. By changing the axial couplings of the
quarks, i.e. $g_{1us} \simeq 0.9$, we could
improve the rates of both reactions, but the ratios $g_1/f_1$ clearly
force us to use $g_{1us} =1$. Another modification could be the
$\Lambda-\Sigma^0$-mixing, which was considered in Ref.~\cite{karl94}.
Let us write
\begin{eqnarray}
\Lambda_{\text{phys}} &=& \Lambda \cos\phi + \Sigma^0 \sin\phi\;,
\nonumber \\
\Sigma^0_{\text{phys}} &=& - \Lambda \sin\phi + \Sigma^0 \cos\phi\;.
\end{eqnarray}
A reasonable value for the mixing angle is $\phi = -0.015$~\cite{karl94}
which lies within one standard deviation of experiment~\cite{bour82}. The
decay rate and the ratio $g_1/f_1$ are only modified by some percent with
this mixing angle, not helping the disagreement between theory and
experiment.

This inconsistency
of our values is a general feature of quark models with a SU(6)
flavor-spin symmetry~\cite{bag}.
The ratio
$g_1/f_1$ can generally be written as
\begin{equation}
{g_1 \over f_1} = \rho\eta\left({g_1 \over f_1}\right)_{\text{non-rel}}\;,
\end{equation}
where $(g_1/f_1)_{\text{non-rel}}$ is the non-relativistic value.
The quantity $\rho$ is a relativistic suppression factor due to the
`` small '' components in the quark spinors (in the bag-model) or
due to the Melosh-transformation
(in our model). The quantity $\eta$ is an
enhancing factor due to SU(3) symmetry breaking in $\Delta S = 1$
transitions. From Tables \ref{tab:nstrange} and \ref{tab:strange} we see that
$\rho \simeq 0.73 - 0.76$~\cite{schl94b}
depending on the strangeness content of
the wave functions and $\eta \simeq 1.11$. This simple estimate
shows that every quark model is {\it a priori} constrained to
\begin{equation}
{g_1/f_1(\Lambda\to pe^-\bar\nu_e) \over g_1/f_1(\Sigma^-\to ne^-\bar\nu_e)}
= -3
\label{eq:ratio}
\end{equation}
in contrast to the experimental value $-2.11\pm 0.15$ for $g_2 = 0$.
This puzzle was pointed out independently by Lipkin~\cite{lipkin92} and
the author~\cite{schl92}. For
$g_2 \ne 0$ it is measured that \cite{jens83}
\begin{equation}
\left| {g_1 \over f_1}\right|_{\Lambda p} = 0.715 + 0.28 {g_2 \over f_1}\;,
\label{eq:5.31}
\end{equation}
and \cite{hsue88}
\begin{equation}
\left| {g_1 \over f_1} - 0.237{g_2 \over f_1} \right|_{\Sigma^-n} =
0.34 \pm 0.017\;,
\end{equation}
which will bring the data closer to $-3$, but in our model $g_2/g_1 \simeq
0.025$ which is much too small to remove the discrepancy.

\subsection{Configuration mixing}

In this Section we investigate the effect caused by configuration mixing
suggested by spectroscopy. The analysis of the $\Delta$-nucleon mass
splitting suggests~\cite{isgu82,caps86}:
\begin{equation}
| \text{Baryon} \rangle = \text{A}\; [56,0^+] + \text{B}\;
[56,0^+]^* + \text{C}\; [70,0^+],\label{eq:mixing}
\end{equation}
in the notation [SU(6),L$^p$],
where
$
\text{A}^2+\text{B}^2+\text{C}^2=1\;,
$
L denotes the angular momentum, and $p$ is the parity of the nucleon.
The values for A, B, C are listed in Table \ref{tab:mixing} for different
references.

Unfortunately, the mixing configuration does not improve the fit, it
is even worse for the crucial ratio in Eq.~(\ref{eq:ratio}). A rough
estimate gives
\begin{equation}
{g_1/f_1(\Lambda\to pe^-\bar\nu_e) \over g_1/f_1(\Sigma^-\to ne^-\bar\nu_e)}
\simeq -3\left( 1+ \frac{8}{3}\text{C}^2\right) = -3.5\pm 0.1\;,
\label{eq:c2}
\end{equation}
to be compared with the value $-3$ for no mixing,
and the experimental data $-2.11\pm 0.15$.
Other values like the ratio $\mu (p)/\mu (n)$ also get worse with the
configuration mixing suggested in Eq.~\ref{eq:mixing}.
A configuration mixing has recently been suggested in the context
of deep inelastic scattering~\cite{dzie94}. Equation~\ref{eq:c2}
shows that such a possibility is not favored for hyperon decays.

\subsection{The form factors $f_2(0)$ and $g_2(0)$}
Our model agrees with the conserved vector current (CVC)
 hypothesis.
The deviations have the same origin as the too small neutron magnetic
moment~\cite{schl93a}
since $f_2$ and the magnetic moments have similar
analytic forms. The
experimental situation is not yet clear,
some experiments favor \cite{hsue88}
and some disfavor \cite{dwor90} CVC.

For $\Delta S = 1$ transitions the prediction of $g_2/g_1$ for
nonrelativistic quark models is $\sim$~0.37 and for the bag model
 $\sim$~0.15
\cite{dono82}. Our model gives also a constant value
\begin{equation}
\left({g_2 \over g_1}\right)_{\Delta S =1}\simeq 0.025\;.
\end{equation}
For $\Delta S =0$ transitions we get
\begin{equation}
\left({g_2 \over g_1}\right)_{\Delta S =0}\simeq 0.0033\;,
\end{equation}
if we put $m_d-m_u = 7$ MeV.
This confirms the viewpoint of the PDG \cite{part94} which fixes $g_2=0$.
Experiments also find a vanishing or small $g_2$ \cite{garc85}.

With CVC and the absence of $g_2$ we reach the same conclusion that was
reached in nuclear physics.

\subsection{$K^2$-dependence of the form factors}
Tables \ref{tab:nstrange} and \ref{tab:strange} suggest that the form
factor of Eq.~(\ref{eq:pol}) can be approximated by the dipole form
\begin{equation}
f(K^2) \simeq {f(0) \over \left( 1-K^2 / \Lambda_2^2\right)^2} \;.
\end{equation}
The axial vector form factor $g_1$ for the neutron decay
 gives a
value $M_A = \Lambda_2 = 1.04$ GeV compared to the experimental
value $M_A = (1.00 \pm 0.04)$ GeV \cite{beli85,brun90}.

If we take the dipole Ansatz we can compare our values for $M_V$ and $M_A$
with the results of other work (see Table \ref{tab:dipole}).

The contribution of $M_V$ and $M_A$ to the rate and to $x = g_1/f_1$
to first order is
\begin{eqnarray}
{\Delta \Gamma \over \Gamma} &=& {8 \over 7}{\beta^2M^2 \over (1+3x^2)}
\left({1 \over M_V^2}+{5x^2 \over M_A^2}\right)\nonumber\\
&&\\
{\Delta x^2 \over x^2} &=&-{8 \over 7}\beta^2M^2\left[{(1 -
\alpha_{e\nu})\alpha_{e\nu}
 \over M_V^2} + {6+5\alpha_{e\nu} \over M_A^2}\right] \nonumber
\end{eqnarray}
which shows that our parameters give for the decay
$\Sigma^-\to n e^-\bar\nu_e$ a 0.3\% larger rate and a
4\% smaller $g_1/f_1$ than with the parameters of Gaillard
{\it et al.} that
are often used for the experimental analysis. Although this does not
explain the inconsistency of the data with our calculation, it shows that
future high-statistics experiments should pay more attention to
$M_V$ and $M_A$ in analyzing $g_1/f_1$.

\subsection{SU(3) symmetry breaking}

There are some questions concerning flavor SU(3) breaking in
semileptonic weak hyperon decays~\cite{roos90,ratc90,gens90}.
In a recent, careful analysis Ref.~\cite{gens93} shows that there is both
consistency and evidence for SU(3) breaking.
The SU(3) symmetry breaking for $f_1$ and $g_1$ within our model is given
in Tables~\ref{tab:XI} and \ref{tab:XII}, respectively. It originates
from the mass difference $\Delta m=m_s-m_{u/d}$; and it is included to
all orders of $\Delta m$ in our approach. The values in the present model
are similar to the bag model calculation of Ref.~\cite{dono82}. Note that
the center of mass corrections are already included in our formalism.
Reference~\cite{garc92} suggests that $f_1/f_1^{SU(3)} > 1$ to reconcile
the value for $V_{us}$ for both the $K_{l3}$ and hyperon decays. In our
approach we find $f_1/f_1^{SU(3)} < 1$ since the wave function overlap is
smaller for $\Delta m \neq 0$.

In order to determine the Kobayashi-Maskawa-Cabibbo matrix element
$V_{us}$ we can fit the hyperon decay rate and asymmetries within the
Cabibbo model using the $f_1$ and $g_1$ from Tables~\ref{tab:XI} and
\ref{tab:XII}, and using the dipole masses from Table~\ref{tab:dipole}.
We get a value similar to Ref.~\cite{garc92}
\begin{equation}
V_{us}=0.225 \pm 0.003 \cite{schl92}.
\end{equation}
This has to be compared to the value from $K_{e3}$ which is $0.2196\pm
0.0023$~\cite{leut84}. A discussion about this discrepancy can be found
in Ref.~\cite{garc92}.
Note that the matrix element $V_{us}$ is a crucial input for the
determination of all parameters of the CKM matrix in the
framework proposed in Ref.~\cite{bura94}.

\section{Conclusions}\label{sec:5}

In this paper we have analyzed in detail the semileptonic beta decay of
the nucleons and hyperons within a relativistic constituent quark model.
All parameters of the model have previously been determined by a fit to
the magnetic moments of the baryon octet. We see no evidence for
configuration mixing. The momentum dependence of the form factors has been
calculated and we find some deviation from popular parameterizations.
The SU(3) symmetry breaking for the vector and axial form factors is
determined. We find that the symmetry breaking for $g_1(\Lambda \to p)$ is
of second order. Our value for $V_{us}$ is somehow larger than the
$K_{e3}$ one in agreement with other studies~\cite{gens93,garc92}.
We conclude that our relativistic constituent quark model
does a good job in analyzing the electroweak properties of the baryon octet.

\acknowledgments

It is a pleasure to thank S.~J.~Brodsky and F.~Coester for helpful
discussions.
This work was supported in part by the Schweizerischer Nationalfonds and
in part by the Department of Energy, contract DE-AC03-76SF00515.

\narrowtext

\begin{table}
\caption{Parameters in Eq.~(\protect\ref{eq:5.20}).}
\begin{tabular}{cccc}
Reaction& $A(f_1)$ & $A(f_2)$ & $A(g_1)$ \\
\tableline
$np$ & 1 &$(2A_2-5A_3)/3$ & ${5 \over 3} $\\
$\Sigma^+\Lambda$ & 0 &$(A_2+A_1-2A_3)/\sqrt 6$&$ \sqrt{{2 \over 3}}$\\
$\Sigma^-\Lambda$ & 0 &$(A_2+A_1-2A_3)/\sqrt 6$&$ \sqrt{{2 \over 3}}$\\
$\Sigma^-\Sigma^0$ & $\sqrt{2} $&$-(4A_3+A_2+A_1)/(3\sqrt 2)$
&$ {2\sqrt{2} \over 3}$\\
$\Sigma^0\Sigma^+$ &$ -\sqrt{2}$ &$(4A_3+A_2+A_1)/(3\sqrt 2)$
&$-{2\sqrt{2} \over 3}$\\
$\Xi^-\Xi^0$ & --1 &$(2A_2+2A_1-A_3)/3$  &$ {1 \over 3}$\\
\end{tabular}
\label{tab:reaction1}
\end{table}

\begin{table}
\caption{Parameters in Eq.~(\protect\ref{eq:5.22}).}
\label{tab:reaction2}
\begin{tabular}{ccccccc}
Reaction& $B_1$ & $B_2$ & $B_3$ & $B_4$ & $B_5$ & $B_6$ \\
\tableline
$\Lambda p$ &$ -\sqrt{{3 \over 2}}$ & $-\sqrt{{3 \over 2}}$ & 0 &$
-\sqrt{{3 \over 2}}$& 0 & 0 \\
$\Sigma^0 p$ &$ -{1 \over \sqrt{2}}$ &$ {1 \over 3\sqrt{2}}$ &
${\sqrt{2} \over 3}$ & ${1 \over 3\sqrt{2}}$ &$ {4\sqrt{2} \over 3}$ &$
{\sqrt{2} \over 3}$ \\
$\Sigma^- n $& --1 &$ {1 \over 3}$ & ${2 \over 3}$ & ${1 \over 3}$ &
${8 \over 3}$ &$ {2 \over 3}$ \\
$\Xi^-\Lambda$ & $\sqrt{{3 \over 2}} $&$ {1 \over \sqrt{6}}$ &$ -{1 \over
\sqrt{6}} $&$ {1 \over \sqrt{6}}$ & $-2\sqrt{{2 \over 3}} $&$ -{1 \over 6}$ \\
$\Xi^-\Sigma^0$ &$ {1 \over \sqrt{2}}$ &$ {5 \over 3\sqrt{2}}$ & ${1 \over
3\sqrt{2}}$ & ${5 \over 3\sqrt{2}} $&$ {4 \over 3\sqrt{2}}$ &$
{\sqrt{2} \over 6}$ \\
$\Xi^0\Sigma^+$ & 1 & ${5 \over 3} $&$ {1 \over 3} $&$ {5 \over 3}$ & $
{4 \over 3}$ &$ {1 \over 3} $\\
\end{tabular}
\end{table}

\mediumtext
\begin{table}
\caption[Results for $\Delta S = 0$ weak beta decay.]
{Results for $\Delta S = 0$ weak beta decay.
Experimental data are from PDG \protect\cite{part94}.}
\label{tab:nstrange}
\begin{tabular}{cccccccc}
 & & $np$ & $\Sigma^+\Lambda $& $\Sigma^-\Lambda $& $\Sigma^-\Sigma^0$ &
$\Sigma^0\Sigma^+$ & $\Xi^-\Xi^0$ \\
\tableline
$f_1$ &$ f_1(0)$ & 1.00 &0 &0& 1.41&--1.41&--1.00\\
&$ \Lambda_1$ (GeV) &0.69&--0.32\tablenotemark[1]&--0.32\tablenotemark[1]
&0.60&0.60&0.56\\
&$ \Lambda_2$ (GeV) &0.96&--1.72\tablenotemark[1]&--1.72\tablenotemark[1]
&0.81&0.81&0.71\\
&&&&&&&\\
$g_1$ &$ g_1(0)$ & 1.25&0.60&0.60&0.69&--0.69&0.24\\
& $\Lambda_1$ (GeV) &0.76&0.77&0.77&0.77&0.77&0.76\\
& $\Lambda_2$ (GeV) &1.04&1.05&1.05&1.04&1.04&1.04\\
&&&&&&&\\
$g_1/f_1 $& Theor. &1.252&0.736\tablenotemark[2]&0.736\tablenotemark[2]
&0.491&0.491&--0.244\\
& Expt. &1.2573&0.742\tablenotemark[2]&--&--&--&$<2\times 10^3$\\
&&$\pm 0.0028$&$\pm 0.018$&&&&\\
&&&&&&&\\
${f_2 \over M} $ (GeV$^{-1}$)& Theor. &1.81&1.04&1.04&0.76&--0.76&0.73\\
& CVC &1.85&1.17&1.17&0.60&--0.60&1.00\\
&&&&&&&\\
${g_2 \over M}$ (GeV$^{-1}$) & &0&0&0&0&0&0\\
&&&&&&&\\
Rate ($10^6 s^{-1}$)  & Theor. &$1.152\times 10^{-9}$&0.24&0.389&1.47
\tablenotemark[3]&3.65\tablenotemark[4]&1.55\tablenotemark[3]\\
$e$-mode & Expt. &1.127$\times 10^{-9}$&0.25&0.387&--&--&--\\
&&$\pm 0.003$&$\pm 0.06$&$\pm 0.018$&&&\\
&&&&&&&\\
$\alpha_{e\nu}$ & Theor. &--0.101&--0.404&--0.412&0.436&0.438&0.793\\
& Expt. &--0.102&--0.35&--0.404&&&\\
&&$\pm 0.005$&$\pm 0.15$&$\pm 0.044$&&&\\
$\alpha_{e}$ & Theor. &--0.112&--0.701&--0.704&0.287&0.288&--0.514\\
& Expt. &--0.1127&&&&&\\
&&$\pm 0.0011$&&&&&\\
$\alpha_{\nu}$ & Theor. &0.989&0.647&0.645&0.850&0.850&--0.314\\
& Expt. &0.997&&&&&\\
&&$\pm 0.028$&&&&&\\
$\alpha_{B}$ & Theor. &--0.548&0.070&0.077&--0.710&--0.711&0.518\\
& Expt. &&&&&&\\
\end{tabular}
\tablenotetext[1]{Instead of $\Lambda_i$ we list $f_1^{(i)}$.}
\tablenotetext[2]{Instead of $g_1/f_1$ we list $\sqrt{3/2} g_1$.}
\tablenotetext[3]{$\times 10^{-6}$.}
\tablenotetext[4]{$\times 10^{-8}$.}
\end{table}

\begin{table}
\caption[Results for $\Delta S = 1$ weak beta decay.]
{Results for $\Delta S = 1$ weak beta decay.
Experimental data are from PDG \protect\cite{part94}.}
\label{tab:strange}
\begin{tabular}{cccccccc}
 & & $\Lambda p $& $\Sigma^0 p $& $\Sigma^- n $& $\Xi^-\Lambda $&
$\Xi^-\Sigma^0 $& $\Xi^0\Sigma^+ $ \\
\tableline
$f_1$ &$ f_1(0)$ & --1.19&--0.69&--0.97&1.19&0.69&0.98\\
&$ \Lambda_1$ (GeV) &0.71&0.64&0.64&0.68&0.75&0.75\\
&$ \Lambda_2 $(GeV) &0.98&0.84&0.90&0.89&1.05&1.05\\
&&&&&&&\\
$g_1$ &$ g_1(0)$ & --0.99&0.19&0.27&0.33&0.94&1.33\\
& $\Lambda_1$ (GeV) &0.81&0.83&0.83&0.81&0.81&0.81\\
& $\Lambda_2$ (GeV) &1.12&1.16&1.16&1.10&1.12&1.12\\
&&&&&&&\\
$g_1/f_1 $& Theor. &0.826&--0.275&--0.275&0.272&1.362&1.362\\
& Expt. &0.718&--&--0.340&0.25&1.287&$< 2.93$\\
&&$\pm 0.015$&&$\pm 0.017$&$\pm 0.05$&$\pm 0.158$&\\
&&&&&&&\\
${f_2 \over M} $(GeV$^{-1}$)& Theor. &--0.85&0.44&0.62&0.070&0.98&1.38\\
& CVC &--1.19&--&1.12&--0.080&1.38&1.95\\
&&&&&&&\\
${g_2 \over M}$ (GeV$^{-1}$) & &--0.025&0.0043&0.0061&--\tablenotemark[1]
&--\tablenotemark[1]&--\tablenotemark[1]\\
&&&&&&&\\
Rate ($10^6 s^{-1}$) & Theor. &3.51&2.72&5.74&2.96&0.549&0.942\\
$e$-mode & Expt. &3.170&--&6.88&3.36&0.53&--\\
&&$\pm 0.058$&&$\pm 0.26$&$\pm 0.19$&$\pm 0.10$&\\
&&&&&&&\\
Rate ($10^6 s^{-1}$) & Theor. &0.58&1.18&2.54&0.80&$7.47\times 10^{-3}$&
$7.74\times 10^{-3}$\\
$\mu$-mode & Expt. &0.60&--&3.04&2.1&--&--\\
&&$\pm 0.13$&&$\pm 0.27$&$\pm 2.1$&&\\
&&&&&&&\\
$\alpha_{e\nu}$ & Theor. &--0.100&0.443&0.437&0.531&--0.252&--0.248\\
& Expt. &--0.019&&0.279&0.53&&\\
&&$\pm 0.013$&&$\pm 0.026$&$\pm 0.1$&&\\
$\alpha_{e}$ & Theor. &--0.021&--0.536&--0.537&0.236&--0.226&--0.223\\
& Expt. &0.125&&--0.519\tablenotemark[2]&&&\\
&&$\pm 0.066$&&$\pm 0.104$&&&\\
$\alpha_{\nu}$ & Theor. &0.992&--0.318&--0.318&0.592&0.973&0.973\\
& Expt. &0.821&&--0.230\tablenotemark[2]&&&\\
&&$\pm 0.066$&&$\pm 0.061$&&&\\
$\alpha_{B}$ & Theor. &--0.582&0.568&0.569&--0.519&--0.437&--0.439\\
& Expt. &--0.508&&0.509\tablenotemark[2]&&&\\
&&$\pm 0.065$&&$\pm 0.102$&&&\\
\end{tabular}
\tablenotetext[1]{${g_2 \over g_1M} \simeq 0.023$ since ${g_2 \over
g_1}\simeq$ constant.}
\tablenotetext[2]{From Ref. \cite{hsue88}.}
\end{table}
\narrowtext

\begin{table}
\caption[Parameters for the configuration mixing of the baryon octet.]
{Parameters for the configuration mixing of the baryon octet
given in Eq. (\protect\ref{eq:mixing}) for two different references.}
\begin{tabular}{cccc}
&A&B&C\\
\tableline
Ref. \cite{isgu82}&0.93&--0.29&--0.23\\
Ref. \cite{caps86}&0.90&--0.34&--0.27\\
\end{tabular}
\label{tab:mixing}
\end{table}

\mediumtext
\begin{table}
\caption{The parameters $M_V$ and $M_A$ for various models in units of GeV.}
\label{tab:dipole}
\begin{tabular}{ccccccccc}
&\multicolumn{2}{c}{This work}&\multicolumn{2}{c}{Gaillard {\it et al.}
\cite{gail84}}&\multicolumn{2}{c}{Garcia {\it et al.} \cite{garc85}}
&\multicolumn{2}{c}{Gensini \cite{gens90}}\\
&$M_V$&$M_A$&$M_V$&$M_A$&$M_V$&$M_A$&$M_V$&$M_A$\\
\tableline
$np$&0.96&1.04&0.84&1.08&0.84&0.96&0.84&1.08\\
$\Sigma\Lambda$&-&1.05&-&1.08&-&0.96&-&1.08\\
$\Sigma\Sigma$&0.81&1.04&0.84&1.08&0.84&0.96&0.84&1.08\\
$\Xi\Xi$&0.71&1.04&0.84&1.08&0.84&0.96&0.84&1.08\\
$\Lambda p$&0.98&1.12&0.98&1.25&0.97&1.11&0.94&1.16\\
$\Sigma p$&0.84&1.16&0.98&1.25&0.97&1.11&0.94&1.16\\
$\Sigma n$&0.90&1.16&0.98&1.25&0.97&1.11&0.94&1.16\\
$\Xi\Lambda$&0.89&1.10&0.98&1.25&0.97&1.11&0.94&1.16\\
$\Xi\Sigma$&1.05&1.12&0.98&1.25&0.97&1.11&0.94&1.16\\
\end{tabular}
\end{table}
\narrowtext

\begin{table}
\caption{Symmetry breaking for $f_1$. The ratio $f_1/f_1^{\text{SU(3)}}$
is shown.}
\label{tab:breakf}
\begin{tabular}{ccccc}
&This work&Donoghue \cite{dono82} & Krause \cite{krau90} & A\&L
\cite{ande93} \\
\tableline
$\Delta S = 0$&1.000&1.000&1.000&1.000\\
$\Lambda p$&0.976&0.987&0.943&1.024\\
$\Sigma p$&0.975&0.987&-&-\\
$\Sigma n$&0.975&0.987&0.987&1.100\\
$\Xi\Lambda$&0.976&0.987&0.957&1.059\\
$\Xi\Sigma$&0.976&0.987&0.943&1.011\\
\end{tabular}
\label{tab:XI}
\end{table}

\begin{table}
\caption{Symmetry breaking for $g_1$. The ratio $g_1/g_1^{\text{SU(3)}}$
is shown.}
\label{tab:breakg}
\begin{tabular}{ccc}
&This work&Donoghue \cite{dono82} \\
\tableline
$np$&1.000&1.000\\
$\Sigma\Lambda$&0.981&0.9383/0.9390\\
$\Sigma\Sigma$&0.982&-\\
$\Xi\Xi$&0.977&-\\
$\Lambda p$&1.072&1.050\\
$\Sigma p$&1.051&-\\
$\Sigma n$&1.056&1.040\\
$\Xi\Lambda$&1.072&1.003\\
$\Xi\Sigma$&1.061&0.9954\\
\end{tabular}
\label{tab:XII}
\end{table}

\end{document}